\documentstyle[prd,aps]{revtex}

\newcommand{\bea}{\begin{eqnarray}}
\newcommand{\eea}{\end{eqnarray}}

\begin{document}

%%%%%%%%%%%%%%%%%%%%%%%%%%%%%%%%%%%%%%%%%%%%%%%%%%%%%%%%%%%%%%%
\draft
%  For 2 column format.
\twocolumn[\hsize\textwidth\columnwidth\hsize\csname
@twocolumnfalse\endcsname
%%%%%%%%%%%%%%%%%%%%%%%%%%%%%%%%%%%%%%%%%%%%%%%%%%%%%%%%%%%%%%%

\title{Unified Analysis of Cosmological Perturbations in Generalized Gravity}
\author{Jai-chan Hwang}
\address{Department of Astronomy and Atmospheric Sciences,
         Kyungpook National University, Taegu, Korea}
\date{\today}
\maketitle

%%%%%%%%%%%%%%%%%%%%%%%%%%%%%%%%%%%%%%%%%%%%%%%%%%%%%%%%%%%%%%%
\begin{abstract}
In a class of generalized Einstein's gravity theories we
derive the equations and general asymptotic solutions
describing the evolution of the perturbed universe in unified forms.
Our gravity theory considers general couplings between the scalar field
and the scalar curvature in the Lagrangian, thus includes broad classes of
generalized gravity theories resulting from recent attempts for the
unification.
We analyze both the scalar-type mode and the gravitational wave
in analogous ways.
For both modes the large scale evolutions are characterized by
the same conserved quantities which are valid in the Einstein's gravity.
This unified and simple treatment is possible due to our proper choice
of the gauges, or equivalently gauge invariant combinations.
\end{abstract}

%%%%%%%%%%%%%%%%%%%%%%%%%%%%%%%%%%%%%%%%%%%%%%%%%%%%%%%%%%%%%%%
%  For 2 column format.
\vskip2pc]
%%%%%%%%%%%%%%%%%%%%%%%%%%%%%%%%%%%%%%%%%%%%%%%%%%%%%%%%%%%%%%%

%%%%%%%%%%%%%%%%%%%%%%%%%%%%%%%%%%%%%%%%%%%%%%%%%%%%%%%%%%%%%%%
Recently, we notice a remarkable growth of interest concerning the roles of
generalized Einstein's gravity theories, particularly, in the contexts of
the unification, quantum gravity and cosmology.
Low energy limits of diverse attempts to unify the gravity with other
fundamental forces show some modified gravity theories involving
dilaton field as the natural outcome; examples are superstring theories
and theories implementing Kaluza-Klein idea \cite{Alvarez,Cho}.
Quantum loop corrections in the context of quantum fields in curved spacetime
and arguments for renormalization favor the gravity with higher order
curvature terms and general couplings with scalar field \cite{QFCS}.
Consequently, there exists growing possibility that the dynamics of the early
universe was governed by a gravity theory more general
than the Einstein's one.
In fact, it is known that some classes of generalized gravity naturally
provide the inflation stage in the early universe \cite{Starobinsky}.

In the following we present a unified way of treating the cosmological
perturbations in a class of generalized gravity theories.
The equations and general asymptotic solutions for the scalar-type mode
and the gravitational wave are derived in unified forms.
Remarkably, neglecting the transient mode, the simple forms of the large
scale solutions known for a minimally coupled scalar field in the
Einstein's gravity, remain valid in our generalized gravity theories,
(\ref{UCG-delta-phi-LS-sol},\ref{GW-sol}).

We consider a class of generalized gravity theories represented by an action
\bea
   & & S = \int d^4 x \sqrt{-g} \left[ {1 \over 2} f (\phi, R)
       - {1\over 2} \omega (\phi) \phi^{;a} \phi_{,a} - V(\phi) \right],
   \label{Action}
\eea
where $\phi$ and $R$ are the scalar field and the scalar curvature,
respectively.
We call it generalized $f(\phi, R)$ gravity.
We note that $f$ is a general function of $\phi$ and $R$, and
$\omega$ and $V$ are general functions of $\phi$.
The gravitational field equation and the equation of motion are:
\bea
   & & G_{ab} = {1\over F} \Bigg[
       \omega \left( \phi_{,a} \phi_{,b}
       - {1\over 2} g_{ab} \phi^{;c} \phi_{,c} \right)
   \nonumber \\
   & & \qquad \qquad
       - g_{ab} {RF - f + 2 V \over 2} + F_{,a;b}
       - g_{ab} {F^{;c}}_{c} \Bigg],
   \label{GGT-GFE} \\
   & & {\phi^{;a}}_{a} + {1 \over 2 \omega} \left(
       \omega_{,\phi} \phi^{;a} \phi_{,a}
       + f_{,\phi} - 2 V_{,\phi} \right) = 0,
   \label{GGT-EOM}
\eea
where $F \equiv \partial f/ \partial R$.

The generalized $f(\phi,R)$ gravity theory in (\ref{Action})
includes the following gravity theories as cases:
(A) $f(R)$ gravity: $f=f(R)$ and $\phi = 0$.
$R^2$ gravity is a case with $f(R) = R - R^2 / (6 M^2)$.
(B) Generalized scalar tensor theories:
$f = 2 \phi R$ and $\omega \rightarrow 2 \omega(\phi) / \phi$.
Brans-Dicke theory is a case with $V = 0$ and $\omega = {\rm constant}$.
In the higher dimensional unification the Kaluza-Klein dilaton
plays the role of the Brans-Dicke scalar field with a difference in the
sign of $\omega$ \cite{Cho}.
(C) $F(\phi)R$ gravity: $f = F(\phi) R$.
This includes diverse cases of dilaton coupling.
(D) Generally coupled scalar field:
$f = \left( \gamma - \xi \phi^2 \right) R$ and $\omega = 1$
where $\gamma$ and $\xi$ are constants.
The nonminimally coupled scalar field is a case with $\gamma = 1$;
a minimally coupled scalar field is a case with $\xi = 0$.
Induced gravity is a case with $\gamma = 0$ and a special potential.

As a background model universe we take a spatially homogeneous and isotropic
metric (FLRW model) with vanishing spatial curvature and cosmological constant.
The most general scalar-type perturbations of the FLRW spacetime can be
written as
\bea
   d s^2
   &=& - \left( 1 + 2 \alpha \right) d t^2
       - \chi_{,\alpha} d t d x^\alpha
   \nonumber \\
   & &
       + a^2 \delta_{\alpha\beta}
       \left( 1 + 2 \varphi \right) d x^\alpha d x^\beta,
   \label{metric-scalar}
\eea
where $a(t)$ is the cosmic scale factor.
Without losing generality we took a unique spatial gauge condition which
removes the spatial gauge mode completely; this made the spatial part of the
metric of (\ref{metric-scalar}) simple (see \S 3 of \cite{MSF-UCG}).
For the scalar field we let
$\phi ({\bf x}, t) = \phi (t) + \delta \phi ({\bf x}, t)$;
similarly we use $F ({\bf x}, t) = F (t) + \delta F ({\bf x}, t)$.
The perturbed order quantities $\alpha ({\bf x}, t)$, $\varphi ({\bf x}, t)$,
$\chi ({\bf x}, t)$, and $\delta \phi ({\bf x}, t)$ are spatially
gauge-invariant, but are temporally gauge dependent; letting any one of
these variables equal to zero can be used as a gauge condition.
Besides the scalar-type mode there exist two other types of fluctuations
(the rotation and the gravitational wave) which decouple
from the scalar-type one in the FLRW background.
In the theories we are considering in (\ref{Action})
$\delta \phi$ (or $\delta F$)
does not directly couple with these other types of fluctuations \cite{GGT1}.

In the context of cosmological perturbations
we proposed a practical way of analysing them which
we will call a gauge ready method \cite{Bardeen1988,PRW}.
Using this method, having the gauge freedom can be used as an advantage
in analysing problems.
We first write equations without imposing the temporal gauge condition;
in a homogeneous and isotropic background the spatial gauge condition becomes
trivial.
We can choose the temporal gauge condition based on the consequent
mathematical simplification to handle the problem.
In \cite{PRW} we found several temporal gauge conditions which fix the
temporal gauge transformation property completely.
Any variable under such gauge conditions can be written in terms of
the corresponding unique gauge invariant combination;
thus, we can regard them as equivalently gauge invariant ones,
see (\ref{UCG-UFG}).
Having a solution in a given gauge condition, the rest of the solutions
even in other gauge conditions can be derived easily.
The gauge ready formulation provides a simple and systematic
way for translating solutions between different gauge conditions.

In the uniform-curvature gauge, we take $\varphi \equiv 0$ as
the gauge condition \cite{PRW}.
In order to derive a closed form equation for $\delta \phi$ (or $\delta F$)
we can use the following gauge invariant combination
\bea
   & & \delta \phi_\varphi
       \equiv \delta \phi - {\dot \phi \over H} \varphi
       \equiv - {\dot \phi \over H} \varphi_{\delta \phi},
   \label{UCG-UFG}
\eea
where $H \equiv \dot a / a$;
for gauge transformation properties, see \S 2.2 of \cite{PRW}.
(\ref{UCG-UFG}) provides a relation between the perturbed curvature
variable ($\varphi$) in the uniform-field gauge and perturbed field
variable ($\delta \phi$) in the uniform-curvature gauge;
$\delta \phi_\varphi$ becomes $\delta \phi$ in the uniform-curvature gauge,
and $\varphi_{\delta \phi}$ becomes $\varphi$ in the uniform-field gauge
which takes $\delta \phi \equiv 0$ as the gauge condition.
In the specific generalized gravity theories in (A)-(D)
$\delta F$ and $\delta \phi$ are related to each other as
\bea
   & & {\delta F \over \dot F} = {\delta \phi \over \dot \phi}.
   \label{F-phi}
\eea
Thus, we will consider $\delta \phi$ as the representative one.
A shortcut, but fully rigorous, way for deriving the equation for
$\delta \phi_\varphi$ is obtained by an equation for $\varphi_{\delta \phi}$.
Derivation of the $\varphi_{\delta \phi}$ equation is considerably easier
compared with the one for $\delta \phi_\varphi$.

In the uniform-field gauge we let $\delta \phi = 0$, thus from (\ref{F-phi})
$\delta F = 0$.
In \cite{PRW} a complete set of
equations describing the scalar type perturbations in the
generalized $f(\phi, R)$ gravity is presented in a gauge ready
form.
{}From Eqs.(69-74) of \cite{PRW} we can derive a second order differential
equation for $\varphi_{\delta \phi}$.
After a straightforward algebra we can derive
\bea
   & & { \left( H + {1\over 2} {\dot F \over F} \right)^2 \over
       a^3 \left( \omega \dot \phi^2 + {3 \over 2} {\dot F^2 \over F} \right) }
       \left[ { a^3 \left( \omega \dot \phi^2 + {3 \over 2}
       {\dot F^2 \over F} \right) \over
       \left( H + {1\over 2} {\dot F \over F} \right)^2 }
       \dot \varphi_{\delta \phi} \right]^\cdot
       + {k^2 \over a^2} \varphi_{\delta \phi}
   \nonumber \\
   & & \qquad
       = 0.
   \label{UFG-varphi-eq}
\eea
In the large scale limit, thus ignoring the $k^2 / a^2$
term, we have the following integral form solution
\bea
   & & \delta \phi_\varphi ({\bf x}, t)
       = - {\dot \phi \over H} \varphi_{\delta \phi} ({\bf x}, t)
   \nonumber \\
   & & \qquad
       = - {\dot \phi \over H} \left[ C ({\bf x}) - D ({\bf x}) \int_0^t
       { \left( H + {1\over 2} {\dot F \over F} \right)^2 \over
       a^3 \left( \omega \dot \phi^2 + {3 \over 2} {\dot F^2 \over F} \right) }
       dt \right],
   \label{UCG-delta-phi-LS-sol}
\eea
where $C({\bf x})$ and $D({\bf x})$ are coefficients of the growing
and the decaying modes, respectively.
The growing mode of $\varphi_{\delta \phi}$ is conserved.
We note that in the large scale expansion the decaying mode is higher order
compared with the solutions in the other gauge \cite{Hwang-Minn}.
Thus, the coefficient $C({\bf x})$ can be interpreted as $\varphi_{\delta
\phi}$
in the large scale.

In terms of $\delta \phi_\varphi$, (\ref{UFG-varphi-eq}) can be arranged as
\bea
   \delta \ddot \phi_\varphi
   &+& \Bigg\{ 3 H + { \left( 1 + {\dot F \over 2 H F} \right)^2 \over
       \omega + {3 \dot F^2 \over 2 \dot \phi^2 F } }
       \left[ { \omega + {3 \dot F^2 \over 2 \dot \phi^2 F } \over
       \left( 1 + {\dot F \over 2 H F} \right)^2 } \right]^\cdot \Bigg\}
       \delta \dot \phi_\varphi
   \nonumber \\
   &+& \Bigg\{ {k^2 \over a^2} - {H \over a^3 \dot \phi}
       { \left( 1 + {\dot F \over 2 H F} \right)^2 \over
       \omega + {3 \dot F^2 \over 2 \dot \phi^2 F} }
   \nonumber \\
   & & \quad
       \times
       \left[ { \omega + {3 \dot F^2 \over 2 \dot \phi^2 F} \over
       \left( 1 + {\dot F \over 2 H F} \right)^2 }
       a^3 \left( {\dot \phi \over H} \right)^\cdot \right]^\cdot
       \Bigg\} \delta \phi_\varphi = 0.
   \label{UCG-delta-phi-eq}
\eea
Introducing new variables
\bea
   & & v ({\bf k}, t) \equiv { \sqrt{ \omega
       + {3 \dot F^2 \over 2 \dot \phi^2 F} }
       \over 1 + {\dot F \over 2 H F} } a \delta \phi_\varphi
       = z { H \over \dot \phi } \delta \phi_\varphi
       = - z \varphi_{\delta \phi},
   \nonumber \\
   & &
       z \equiv { \sqrt{ \omega + {3 \dot F^2 \over 2 \dot \phi^2 F} } \over
       1 + {\dot F \over 2 H F} } {a \dot \phi \over H},
   \label{v-def}
\eea
(\ref{UCG-delta-phi-eq}) can be written as
\bea
   & & v^{\prime\prime} + \left( k^2
       - {z^{\prime\prime} \over z} \right) v = 0,
   \label{v-eq}
\eea
where a prime denotes the time derivative based on the conformal time $\eta$,
$d\eta = a^{-1} dt$.

Evolution of (\ref{v-eq}) with different $z$ was studied previously
\cite{v-LS-sol-comment}.
In the small scale limit ($z^{\prime\prime}/z \ll k^2$)
we have $v = c_1 e^{ik\eta} + c_2 e^{-ik\eta}$, thus
\bea
   & & \delta \phi_\varphi ({\bf k}, \eta)
       = {1\over a} { 1 + {\dot F \over 2 H F} \over
       \sqrt{ \omega + {3 \dot F^2 \over 2 \dot \phi^2 F} } }
       \Big[ c_1 ({\bf k}) e^{ik \eta} + c_2 ({\bf k}) e^{-ik \eta} \Big].
   \nonumber \\
   \label{delta-phi-SS-sol}
\eea
For $z^{\prime\prime} / z = n / \eta^2$ with $n = {\rm constant}$, (\ref{v-eq})
becomes a Bessel equation with an exact solution \cite{GGT2,Stewart-Lyth}
\bea
   & & v ({\bf k}, \eta) = \sqrt{|\eta|} \left[
       C_1 ({\bf k}) H_\nu^{(1)} (k|\eta|)
       + C_2 ({\bf k}) H_\nu^{(2)} (k|\eta|) \right],
   \nonumber \\
   & & \qquad
       \nu \equiv \sqrt{ n + {1\over 4} }.
   \label{z-nu-sol}
\eea
We emphasize that (\ref{UFG-varphi-eq}-\ref{z-nu-sol})
are valid for the classes of gravity theories in (A)-(D)
with general $V(\phi)$, $\omega(\phi)$, and $F(\phi)$ (or $F(R)$).
It is noteworthy that (\ref{UFG-varphi-eq}-\ref{z-nu-sol})
express various cases of gravity theories in unified forms.
The equation and the asymptotic solutions for $\delta \phi_\varphi$
are simpler than the ones derived in the zero-shear gauge condition
\cite{ZSG-comment}.

In the case of $f(R)$ gravity without the scalar field
$F(R)$ plays similar role as the scalar field.
(\ref{UFG-varphi-eq}-\ref{z-nu-sol}) are valid for $\delta F$ by replacing
$\delta \phi$ using (\ref{F-phi}); we replace $\delta \phi_\varphi$ with
$(\dot \phi / \dot F) \delta F_\varphi$ and let $\phi = 0$.
(\ref{UCG-delta-phi-eq}) looks cumbersome for such replacement;
a little trick [which can be confirmed by a straight algebra
from (\ref{UFG-varphi-eq})] is that we replace $\delta \phi_\varphi$ with
$\delta F_\varphi$, $\dot \phi$ with $\dot F$, and let $\omega = 0$.

A minimally coupled scalar field in the uniform-curvature gauge was
studied previously \cite{H-QFT,MSF-UCG,H-MSF,Hwang-Minn}.
We have $F = 1$ and $\omega = 1$.
(\ref{UCG-delta-phi-eq}) reduces to
\bea
   & & \delta \ddot \phi_\varphi + 3 H \delta \dot \phi_\varphi
       + \left\{ {k^2 \over a^2} - {H \over a^3 \dot \phi}
       \left[ a^3 \left( {\dot \phi \over H} \right)^\cdot \right]^\cdot
       \right\} \delta \phi_\varphi = 0.
   \nonumber \\
   \label{MSF-eq}
\eea
The solution in (\ref{UCG-delta-phi-LS-sol}) becomes
\bea
   & & \delta \phi_\varphi ({\bf x},t)
       = - {\dot \phi \over H} \left[ C ({\bf x})
       - D ({\bf x}) \int^t_0 {H^2 \over a^3 \dot \phi^2 } dt \right].
   \label{MSF-LS-sol}
\eea
We note that this solution is valid for arbitrary $V(\phi)$, which is true
as long as the background dynamics is governed by the scalar field.
If the background model supported by the scalar field is expanding in power-law
$a \propto t^{2/[3(1+{\rm w})]}$ with
${\rm w} \equiv p/\mu = {\rm constant}$, we have
$\dot \phi/ H = {\rm constant}$ [see Eq.(21) of \cite{H-QFT}];
an exponentially expanding background can be considered as a case of the
power-law case \cite{QFCS-comment}.
Thus, in the power-law case (\ref{v-def}) gives
$z^{\prime\prime}/z = a^{\prime\prime}/a \propto \eta^{-2}$,
and an exact solution is given by (\ref{z-nu-sol}) with
$\nu = {3({\rm w} -1) \over 2 (3 {\rm w} +1)}$;
the exponential expansion is a limiting case with ${\rm w} \rightarrow -1$,
thus $\nu = 3/2$.

The equation for decoupled gravitational waves in the
generalized $f(\phi, R)$ gravity is derived in Eq.(39) of \cite{GGT1}
\bea
   & & \ddot H_T + \left( 3 H + {\dot F \over F} \right) \dot H_T
       + {k^2 \over a^2} H_T = 0.
   \label{GW-eq}
\eea
$H_T$ is gauge invariant; for the notation, see \S 5.2 of \cite{PRW}.
The large scale solution is
\bea
   & & H_T ({\bf x}, t) = C_g ({\bf x})
       - D_g ({\bf x}) \int^t_0 {1\over a^3 F} dt,
   \label{GW-sol}
\eea
where $C_g ({\bf x})$ and $D_g ({\bf x})$ are coefficients of the growing and
decaying modes, respectively.
(\ref{GW-eq},\ref{GW-sol}) for $H_T$ can be compared with
(\ref{UFG-varphi-eq},\ref{UCG-delta-phi-LS-sol}) for $\varphi_{\delta \phi}$;
growing modes of both quantities are conserved in the large scale.
(\ref{GW-eq}) can be transformed into a form similar to (\ref{v-eq}):
\bea
   & & v_g^{\prime\prime} + \left( k^2 - {z_g^{\prime\prime} \over z_g} \right)
       v_g = 0,
   \nonumber \\
   & & v_g \equiv a \sqrt{F} H_T = z_g H_T, \quad
       z_g \equiv a \sqrt{F}.
   \label{vg-eq}
\eea
Similarly as in (\ref{delta-phi-SS-sol}), in the small scale limit we have
\bea
   & & H_T ({\bf k}, \eta) = { 1 \over a \sqrt{F} }
       \Big[ c_{g1} ({\bf k}) e^{ik\eta}
       + c_{g2} ({\bf k}) e^{-ik\eta} \Big].
   \label{GW-SS-sol}
\eea

We comment on a way of managing our perturbation equations and
formal solutions in some concrete cosmological models.
The equations and the general asymptotic solutions derived above
are valid for the classes of the gravity theories mentioned in (A)-(D).
In order to derive the solutions in explicit forms, all we need to know
are the evolution of the background universe characterized by
the scale factor $a(t)$ and probably the scalar field $\phi (t)$.
Equations for the background are presented in Eqs.(68,76) of \cite{PRW}:
\bea
   & & H^2 = {1 \over 3F} \left( {\omega \over 2} \dot \phi^2
       + {RF - f + 2 V \over 2} - 3 H \dot F \right),
   \label{BG-G1} \\
   & & \dot H = - {1\over 2 F} \left( \omega \dot \phi^2
       + \ddot F - H \dot F \right),
   \label{BG-G2} \\
   & & \ddot \phi + 3 H \dot \phi + {1\over 2 \omega}
       \left( \omega_{,\phi} \dot \phi^2 - f_{,\phi} + 2 V_{,\phi}
       \right) = 0.
   \label{BG-G3}
\eea
(\ref{BG-G3}) follows from (\ref{BG-G1},\ref{BG-G2}).
After selecting the favorite gravity theory (thus specifying
$F$ and $\omega$), one can try to find
the evolution of the background by solving (\ref{BG-G1}-\ref{BG-G3}).
Having solutions for the background, the rest are straightforward integration
for the scalar mode using
(\ref{UCG-delta-phi-eq},\ref{UCG-delta-phi-LS-sol},\ref{delta-phi-SS-sol}),
and for the gravitational wave using
(\ref{GW-eq},\ref{GW-sol},\ref{GW-SS-sol}).
{}From a known solution in a gauge, for example $\delta \phi_\varphi$,
we can derive other variables in any gauge as linear combinations
\cite{Hwang-Noh}.

Let us consider a model in which the dynamics in the later stage
of the evolution is governed by
the Einstein gravity.
Evolutions of cosmological perturbations in the context of Einstein gravity
are well known \cite{Lifshitz}.
In the Einstein gravity filled with an ideal fluid,
$- \varphi_\chi$, $v_\chi$, $\delta \mu_v / \mu$, and $\delta T_v / T$
\cite{CG-comment} play the roles of the Newtonian
potential fluctuation ($\delta \Phi$), velocity fluctuation ($\delta v$),
relative density fluctuation ($\delta \varrho / \varrho$),
and the relative temperature fluctuations in the cosmic microwave background
radiation ($\delta T/T$), respectively \cite{H-MDE}.
In the matter dominated stage, ignoring the decaying modes,
we have \cite{H-MDE}:
\bea
   & & \delta \Phi = - {3 \over 5} C, \quad
       \delta v = - {2 \over 5} \left( {k \over aH} \right) C, \quad
       {\delta \varrho \over \varrho} = {2 \over 5}
       \left( {k \over a H} \right)^2 C,
   \nonumber \\
   & &
       {\delta T \over T} = {1\over 5} C.
   \label{pert-sols}
\eea
The coefficient of the growing mode, $C({\bf x})$, is matched so that
it is the same as the one in (\ref{UCG-delta-phi-LS-sol}).
$C({\bf x})$ encodes the spatial structure of the growing mode,
and determines the variables characterizing the perturbed
evolution in the linear regime.
Through the linear evolution the spatial structures are preserved.

If the early universe was governed by a generalized gravity, $C ({\bf x})$
can be determined from the fluctuations in the scalar field using
(\ref{UCG-delta-phi-LS-sol}); we are considering a simple situation where
the scales we consider were always in the large scale limit \cite{MSF-comment}.
Remarkably, for the growing mode, exactly the same solution
known for the Einstein's gravity
remains valid for the generalized gravity (A)-(D);
from (\ref{UCG-delta-phi-LS-sol}), we have
\bea
   & & C({\bf x}) = - {H \over \dot \phi} \delta \phi_\varphi ({\bf x}, t),
\eea
which does not involve $F$ and $\omega$.
[For the gravitational wave, see (\ref{GW-sol}).]
As mentioned before, $C ({\bf x})$ characterizes the spatial structure
of the growing mode of every variable.
According to (\ref{UCG-delta-phi-LS-sol}), $C({\bf x})$ can be interpreted as
a curvature inhomogeneity in the uniform-field gauge
(we write the gravitational wave case together):
\bea
   & & \varphi_{\delta \phi} ({\bf x}, t) = C ({\bf x}), \quad
       H_T ({\bf x}, t) = C_g ({\bf x}),
\eea
which are valid in the corresponding large scale limits.

We note that the simple presentation above is possible due to
our suitable choices of gauge, or equivalently gauge invariant combinations.
The simple and unified forms of equations and asymptotic solutions
derived above will be useful for rigorous treatment of the evolution of
perturbed universe in the context of generalized gravity.

We thank Dr. H. Noh for many useful comments.

%%%%%%%%%%%%%%%%%%%%%%%%%%%%%%%%%%%%%%%%%%%%%%%%%%%%%%%%%%%%%%

%%%%%%%%%%%%%%%%%%%%%%%%%%%%%%%%%%%%%%%%%%%%%%%%%%%%%%%%%%%%%%

\begin{references}
\bibitem{Alvarez}
         E. Alvarez, Rev. Mod. Phys. {\bf 61}, 561 (1989).
\bibitem{Cho}
         Y. M. Cho, Phys. Rev. Lett. {\bf 68}, 3133 (1992).
\bibitem{QFCS}
         N. D. Birrell and P. C. W. Davies, {\it Quantum fields in curved
space}
            (Cambridge, Cambridge University Press, 1982).
\bibitem{Starobinsky}
         A. A. Starobinsky, Phys. Lett. {\bf B91}, 99 (1980);
         D. La and P. J. Steinhardt, Phys. Rev. Lett. {\bf 62}, 376 (1989).
\bibitem{MSF-UCG}
         J. Hwang, Class. Quantum Grav. {\bf 11}, 2305 (1994).
\bibitem{GGT1}
         J. Hwang, Class. Quantum Grav. {\bf 7}, 1613 (1990).
\bibitem{Bardeen1988}
         J. M. Bardeen, in {\it Cosmology and particle physics}, edited by
            L. Fang and A. Zee (London, Gordon and Breach, 1988), 1.
\bibitem{PRW}
         J. Hwang, Astrophys. J. {\bf 375}, 443 (1991).
\bibitem{Hwang-Minn}
         \S 6 of J. Hwang and H. Minn,
            in {\it Proceedings of the second Alexander
            Friedmann international seminar on gravitation and cosmology},
            edited by Yu. N. Gnedin, A. A. Grib and V. M. Mostepanenko
            (St. Petersburg, Friedmann Laboratory Publishing, 1994), 265.
\bibitem{v-LS-sol-comment}
         In the large scale limit ($z^{\prime\prime}/z \gg k^2$) we have
         \cite{Russian,GGT2,Mukhanov-etal}
         $v ({\bf x}, \eta) = c_g ({\bf x}) z + c_d ({\bf x}) z
         \int^\eta_0 d \eta / z^2$.
         By matching $c_g = - C$ and $c_d = D$, it is
         equivalent to (\ref{UCG-delta-phi-LS-sol}).
\bibitem{Russian}
         L. A. Kofman and V. F. Mukhanov, JETP Lett. {\bf 44}, 619 (1986);
         V. F. Mukhanov, L. A. Kofman and D. Yu. Pogosyan,
            Phys. Lett. {\bf B193}, 427 (1987);
         L. A. Kofman, V. F. Mukhanov and D. Yu. Pogosyan,
            Sov. Phys. JETP {\bf 66}, 433 (1988);
         V. F. Mukhanov, Phys. Lett. {\bf B218}, 17 (1989);
         M. B. Baibosunov, V. Ts. Gurovich and U. M. Imanaliev,
            Sov. Phys. JETP {\bf 71}, 636 (1990).
\bibitem{H-GGT}
         J. Hwang, Class. Quantum Grav. {\bf 8}, 195 (1991);
         Class. Quantum Grav. {\bf 8}, L133 (1991);
         \S 4.2 of Astrophys. J. {\bf 380}, 307 (1991).
\bibitem{Mukhanov-etal}
         V. F. Mukhanov, H. A. Feldman and R. H. Brandenberger,
            Phys. Rep. {\bf 215}, {203} (1992).
\bibitem{GGT2}
         J. Hwang, Phys. Rev. D {\bf 42}, 2601 (1990).
\bibitem{Stewart-Lyth}
         E. D. Stewart and D. H. Lyth, Phys. Lett. {\bf B302}, 171 (1993).
\bibitem{ZSG-comment}
         In the zero-shear gauge we let $\chi \equiv 0$;
         $\delta \phi_\chi \equiv \delta \phi - \dot \phi \chi$ is a gauge
         invariant combination which becomes $\delta \phi$ in the
         zero-shear gauge.
         The equation and asymptotic solutions were derived in \cite{GGT2} and
         \S 4.3.2 of \cite{PRW} (see also \cite{H-GGT,Russian,Mukhanov-etal});
         these can also be presented in unified forms \cite{Hwang-Noh}.
\bibitem{Hwang-Noh}
         J. Hwang and H. Noh, unpublished (1995).
\bibitem{H-QFT}
         J. Hwang, Phys. Rev. D {\bf 48}, 3544 (1993).
\bibitem{H-MSF}
         J. Hwang, Astrophys. J. {\bf 427}, 542 (1994);
         Gen. Rel. Grav. {\bf 26}, 299 (1994).
\bibitem{QFCS-comment}
         In these cases (\ref{MSF-eq}) becomes an equation which appears often
         in the context of quantum field in curved spacetime \cite{QFCS}.
         Analyses were made in the cases of power-law \cite{POW} and
         exponential expansion \cite{EXP}.
         Parallel analyses in the context of our perturbative approach were
         made in \cite{H-QFT,MSF-UCG}.
\bibitem{POW}
         L. H. Ford and L. Parker, Phys. Rev. D {\bf 16}, 245 (1977);
         L. F. Abbott and M. B. Wise, Nucl. Phys. {\bf B244}, 541 (1984).
\bibitem{EXP}
         T. S. Bunch and P. C. W. Davies, Proc. R. Soc. {\bf A360}, 117 (1978).
\bibitem{Lifshitz}
         E. M. Lifshitz, J. Phys. (USSR), {\bf 10}, 116 (1946);
         J. M. Bardeen, Phys. Rev. D {\bf 22}, 1882 (1980);
         J. Hwang and J. J. Hyun, Astrophys. J. {\bf 420}, 512 (1994).
\bibitem{CG-comment}
         $\delta \mu_v \equiv \delta \mu - (a/k) \dot \mu v$ and
         $\delta T_v \equiv \delta T - (a/k) \dot T v$ are gauge invariant
         combinations.
         The comoving gauge takes $v \equiv 0$ where
         $v \equiv -(k/a) \Psi/(\mu + p)$;
         for $\Psi$ see \S 2.1.2 of \cite{PRW}.
\bibitem{H-MDE}
         J. Hwang, Astrophys. J. {\bf 427}, 533 (1994);
         Astrophys. J. {\bf 415}, 486 (1993).
\bibitem{MSF-comment}
         For analytic derivations of perturbation spectrums generated
         from quantum fluctuations in the minimally coupled scalar
         field case, see \cite{H-QFT,Stewart-Lyth}.
         Previous attempts in some generalized gravity case
         can be found in \cite{Russian,Mukhanov-etal,pert-GGT-infl}.
         We have not used the conformal transformation in this paper.
\bibitem{pert-GGT-infl}
         D. S. Salopek, J. R. Bond and J. M. Bardeen, Phys. Rev. D {\bf 40},
            1753 (1989);
         E. W. Kolb, D. S. Salopek and M. S. Turner, Phys. Rev. D {\bf 42},
            3925 (1990);
         J. Hwang, Class. Quantum Grav. {\bf 8}, 195 (1991);
         R. Fakir, S. Habib and W. Unruh, Astrophys. J. {\bf 394}, 396 (1992);
         A. H. Guth and B. Jain, Phys. Rev. D {\bf 45}, 426 (1992);
         S. Mollerach and S. Matarrese, Phys. Rev. D {\bf 45}, 1961 (1992);
         N. Deruelle, C. Gundlach and D. Langlois, Phys. Rev. D {\bf 46}, 5337
            (1992);
         M. Gasperini and G. Veneziano, Phys. Rev. D {\bf 50}, 2519 (1994).
\end{references}
\end{document}